\documentclass[preprint,preprintnumbers,amsmath,amssymb]{revtex4}

\begin{document}

\title{A new inflaton model beginning near the Planck epoch}

\author{Reuven Opher  and Ana Pelinson}

\affiliation{Departamento de Astronomia, IAG, Universidade de
S\~{a}o Paulo, 05508-900, S\~{a}o Paulo, SP, Brazil}

\date{\today}

\begin{abstract}
The Starobinsky model predicts a primordial inflation period
without the presence of an inflaton field. The modified version of
this model predicts a simple time dependence for the Hubble
parameter $H(t)$, which decreases slowly between the Planck epoch
and the end of the inflation, $H(t)=M_{\rm Pl}-\beta\,M_{\rm
Pl}^2\,t$, where $\beta$ is a dimensionless constant to be
adjusted from observations.  We investigate an inflaton model
which has the same time dependence for $H(t)$. A reverse
engineered inflaton potential for the time dependence of $H$ is
derived. Normalization of the derived inflaton potential is
determined by the condition that the observed density
fluctuations, $\delta\rho/\rho\approx 10^{-5}$, are created at
$\sim 60\, e$-folds before the end of inflation. The derived
potential indicates an energy (mass) scale, $M_{\rm end}\sim
10^{13}\,{\rm GeV}$, at the end of inflation. Using the slow roll
parameters, which are obtained from this potential, we calculate
the spectral index for the scalar modes $n_S$ and the relative
amplitude of the tensor to scalar modes $r$. A tensor
contribution, $r\simeq 0.13$, and an approximately
Harrison-Zel´dovich density perturbation spectrum, $n_S \simeq
0.95$, are predicted.

\end{abstract}




\maketitle

\section{Introduction}

The two major problems in cosmology are the origins of the
primordial inflation period and the present ``inflation" period of
the universe. It is possible that both origins are linked.
Primordial inflation could have been created by a non-zero vacuum
energy. Subsequently, the vacuum energy could have decayed, creating
the present period of acceleration. However, strong limits were
recently placed on the possible decay of the vacuum energy into cold
dark matter (CDM) or cosmic microwave background (CMB) photons
\cite{braz}.

The most popular model for the origin of primordial inflation
remains the inflaton (scalar field) model. We investigate here an
inflaton model based on the simple time dependence of the Hubble
parameter, $H(t)$ [Eq.(3)], that was predicted by the modified
Starobinsky model \cite{AIM},\cite{AIMmod}. (See \cite{star1} for
the original Starobinsky model.) The Starobinsky model suggests that
quantum fluctuations created a non-zero vacuum energy that induced
the primordial inflation period.

Instead of assuming an ad hoc inflaton potential, as in the standard
inflation model, we use the reverse engineering method of Ellis,
Murugan and Tsagas \cite{ellis1} to derive the inflaton potential
from the $H(t)$ of Eq.(3). The derived potential becomes negligible
at the end of inflation, creating the observed density fluctuations,
$\delta\rho/\rho\approx 10^{-5}$. These fluctuations are determined
by the value of the potential and its first derivative at
$60\,e$-folds before the end of inflation. This condition, together
with the time dependence of the potential, determine a mass (energy)
scale, $M_{\rm end}\simeq 10^{13}{\rm GeV}\sim 10^{-6}M_{\rm Pl}$,
at the end of inflation. From the slow roll parameters obtained from
the derived potential, we calculate the spectral index of the scalar
modes $n_S$ and the relative amplitude of the tensor to scalar modes
$r$. The derived spectral index $n_S$ is in agreement with the WMAP
data \cite{peiris,3WMAP}. The ratio of tensor to scalar modes obtained,
$r\sim 0.13$, is similar to that of most inflation models, which
predict $r\sim 10-30\%$.

We can compare our scale $M_{\rm end}$ at the end of inflation with
the results of Vilenkin \cite{vile2} and Starobinsky \cite{star2}.
Vilenkin noted that, in the Starobinsky model, the Hubble parameter
defines a mass (energy) scale with a limiting value, $M_{\rm
end}\lesssim 10^{16}{\rm GeV}$, at the end of inflation. Starobinsky
predicted that $M_{\rm end}\lesssim 10^{14}{\rm GeV}$ by requiring
that the $\delta\rho/\rho$, resulting from inflation, is
sufficiently small. Our derived value, $M_{\rm end}\sim 10^{13}{\rm
GeV}$, is consistent with the upper limits of both Vilenkin and
Starobinsky for $M_{\rm end}$.

Although the potential that we obtain [Eq.(6)] is superficially
similar to a standard inflation potential that depends on the
square of the massive scalar field (see, for example,
\cite{basset} for a recent review), our inflation model is very
much different from the standard model for the following reasons:

1) The standard massive scalar inflation potential has two free
parameters: the magnitude of the potential and its first derivative
at $\sim 60\,e$-folds before the end of inflation. However, our
potential in Eq.(6) is completely determined by a single parameter
$\beta$, which is derived from the simple time dependence of the
Hubble parameter in Eq.(3);

2) In the standard inflation model, there are many possible forms
that the massive scalar potential can take. However, the form of our
potential, a quadratic dependence on the field, is determined
uniquely by Eq.(3);

3) The origin of the potential in the standard model is completely
unknown. Moreover, there is no clear justification for its form; and

4) In the standard model, the inflation period begins when there is
a displacement of the massive scalar field from the minimum of its
potential. The origin of this displacement is left unexplained and
the epoch in which it occurs is not specified. However, in our
model, the beginning of inflation is specified to occur at the
Planck epoch (i.e., at the beginning of the universe). The origin of
the inflation is a direct result of the simple time dependence of
the Hubble parameter in Eq.(3). Moreover, there is no initial
displacement of the field that is left explained.

We present the algorithm for constructing the potential from the time dependence of
the Hubble parameter in $\S$ 2. In $\S$ 3, we use this algorithm to obtain the
effective potential from the Hubble parameter, $H(t)=M_{\rm Pl}-\beta M_{\rm Pl}^2
t$. From the potential, we calculate the spectral index of the $\delta\rho/\rho$ and
the intensity of primordial gravitational waves. The mass (energy) scale at the end
of inflation, $M_{\rm end}\sim 10^{13}{\rm GeV}$, is determined from the requirement
that the potential creates observed $\delta\rho/\rho\simeq 10^{-5}$ at $\sim
60\,e$-folds before the end of inflation. Finally, our conclusions are presented in
$\S$ 4.


\section{The framework of the single scalar model}

Let us assume that there exists an inflaton field, $\phi=\phi(t)$, where $t$ is the
usual time function, in accordance with the Roberston-Walker symmetry \cite{kolbt}.
The Lagrangian containing a minimally coupled scalar field is
\begin{equation*}
 L=\frac{1}{2}\left(
\partial \phi \right) ^{2}-V\left( \phi
\right)=\frac{1}{2}\, {\dot{\phi}}^{2}- V(\phi) \,,\end{equation*}
where $\dot\phi=d\phi /dt$. The scalar stress tensor takes the perfect fluid form,
\begin{equation*}
T_{ab}=\left( p+\rho \right) u_{a}u_{b}+p\, g_{ab}\,,
\end{equation*}
with the following energy density and pressure of the scalar inflaton field:
\begin{equation*}
\rho _{\phi }=\frac{1}{2}\,{\dot{\phi}}^{2}+V(\phi )\,,
\end{equation*}
\begin{equation*}
p_{\phi }=\frac{1}{2}\,{\dot{\phi}}^{2}-V(\phi ) \,.
\end{equation*}

The classical equation of motion for $\phi(t)$, which follows from
the variation of the action $S=\int d^{4}x\,\sqrt{-g}\,{L}$, is
\begin{equation*}
{\ddot{\phi}}+3H{\dot{\phi}}+\frac{dV}{d\phi }=0 \,,\end{equation*}
where $H={\dot{\sigma}}(t)$. The field equations for the Robertson Walker model,
with $k=0$, are
\begin{equation*}
3\dot{H}+3H^{2}=(8\pi G)(V(\phi
)-\dot{\phi}^{2})\,,
\end{equation*}
\begin{equation*}
3H^{2}=(8\pi G)\left( V(\phi )+\frac{\dot{\phi}^{2}}{2}\right)
\,.\end{equation*}
Following Ellis, Murugan and Tsagas \cite{ellis1}, we combine these
two independent equations to obtain a more convenient set of
equations,
\begin{equation}
V(\phi (t))=\frac{1}{(8\pi G)}\left( \dot{H}+3H^{2}\right)\,,
\label{poteq}
\end{equation}

\begin{equation}
\dot{\phi}^{2}=-\frac{1}{(4\pi G)}\dot{H} \,. \label{phieq}
\end{equation}

From $H(t)$, the above equations have been used to construct the effective
potential in the following manner:\\
\noindent i) Eq.(\ref{phieq}) is integrated to obtain
$\phi(t)$\,;\\
\noindent ii) $t$ as a function of $\phi$ is found; \\
\noindent iii) $t(\phi)$ is
substituted in $H(t)$ to obtain $H(\phi)$\,; and\\
\noindent iv) the potential $V(\phi)$ is obtained, using Eq.(\ref{poteq}).

\section{The effective inflaton potential}

Assuming the simple Hubble parameter time dependence,
\begin{equation}
H(t)=M_{\rm Pl}-\beta M_{\rm Pl}^2 t\,, \label{Hdet}
\end{equation}
we solved Eq.(\ref{phieq}) for $\phi(t)$, obtaining $t$ as a
function of $\phi$,
\begin{equation}
t(\phi)=\pm \frac{1}{M_{\rm Pl}^2 \sqrt{2 \beta}}\left(\phi(t)-\phi_{0}\right),
\label{phidet}
\end{equation}
where $\left| \phi_0 \right|>\left| \phi \right|$. Choosing the
positive sign in Eq.(\ref{phidet}), we have $-\infty < \phi< 0$, as
in \cite{ellis1}. From Eqs.(\ref{Hdet}) and (\ref{phidet}),
\begin{equation}
H(\phi)=M_{\rm Pl} - \sqrt{\frac{\beta}{2}} \left(
\phi(t)-\phi_0\right) \,.
\label{Hphi}
\end{equation}

Following the algorithm of the previous section to obtain $V(\phi)$, we substitute
Eq.(\ref{Hphi}) into Eq.(\ref{poteq}) to obtain
\begin{equation}
V(\phi)=M_{\rm Pl}^4 \left\{-\beta
+3 \left[1-\frac{1}{M_{\rm Pl}}\sqrt{\frac{\beta}{2}} (\phi-\phi_0)\right]^2\right\}
\label{potphie}
\end{equation}
or, in terms of the time,
\begin{equation}
 V(t)=M_{\rm Pl}^2 \left[-\beta M_{\rm Pl}^2 +3\left(M_{\rm Pl}-\beta\,M_{\rm Pl}^2\,
 t\right)^2\right]\,.
 \label{vdet}
 \end{equation}

A realistic potential $V(\phi)$ describing inflation should:\\
\noindent 1) become negligibly small at the end of the inflationary period, so that
there is no important ``cosmological constant" entering
the FRW era; and\\
\noindent 2) produce the density fluctuations at $\sim 60$
$e$-folds before the end of inflation (see e.g. \cite{liddle}),
\begin{equation}
\frac{\delta \rho }{\rho }=\frac{1}{\sqrt{75}\,\pi M_{\rm Pl}^3}
\,\frac{V^{{3}/{2}}(\phi)}{{dV}/{d\phi}} \vert_{N=60}\,,
\label{dflut}
\end{equation}
which are observed to be $\sim 10^{-5}$.

For as long as the first term in Eq.(\ref{Hdet}) dominates, we have the inflationary
expansion $a(t)=\exp{M_{\rm Pl}t}$. The second term in Eq.(\ref{Hdet}) decreases the
expansion rate and is important near the maximum value of $\sigma(t)=\ln{a(t)}$.
Following Vilenkin \cite{vile2}, we characterize the end of inflation by
\begin{equation}
H(t){\vert}_{t=t_{\rm end}}=\mu M_{\rm Pl}\,,
\label{endcond}
\end{equation}
where $H(t_{\rm end})=\dot{\sigma}(t=t_{\rm end})=M_{\rm end}$ and
\begin{equation}
\mu=\frac{M_{\rm end}}{M_{\rm Pl}}
\label{mast}
\end{equation}
is a dimensionless parameter (we should expect $M_{\rm end}< M_{\rm Pl}\sim
10^{19}\,{\rm GeV}$).

The time as a function of $\mu$ at the end of inflation is
\begin{equation}
t_{\rm end}=\frac{1}{\beta M_{\rm Pl}}\left(1-\mu\right) \,.
\label{tend}
\end{equation}

The number of $e$-folds of inflation before $t_{\rm end}$ is
\begin{equation}
N=\int_{t_{60}}^{t_{\rm end}}H(t)dt=\sigma(t_{\rm end})-\sigma(t_{60})\,.
\label{nef}
\end{equation}
We are interested in $N\simeq 60$, the approximate time $t_{60}$, when the observed
$\delta\rho/\rho$ (scalar) and the primordial gravitational (tensor) fluctuations
were created. Substituting Eq.(\ref{tend}) into Eq.(\ref{Hdet}), we find
\begin{equation*}
\sigma_{\rm end}=\sigma(t=t_{\rm end})=\frac{1}{2\beta}\left(1-
\mu^2 \right)\,,
\end{equation*}
where we have used the customary normalization for $a(t=0)=1$. From Eq.(\ref{nef}),
we have
\begin{equation}
\sigma_{60}=\sigma(t=t_{60})=\frac{1}{2\beta}\left(1- \mu^2\right) -60\,.
\label{sigend}
\end{equation}
Using this result to solve Eq.(\ref{Hdet}) for $t_{60}$, we obtain
\begin{equation}
t_{60}=\frac{1}{\beta M_{\rm Pl}}\left[1-\sqrt{1-2\beta\sigma_{60}}\right]
\,.\label{t60}\end{equation}

The slow roll parameters $\epsilon$ and $\eta$ in terms of
$H(\phi)$ are \cite{liddle}
\begin{equation*}\epsilon \equiv 2\,{M_{\rm Pl}^2}\left[
\frac{H^{\,\prime }\left( \phi \right) }{H\left( \phi \right) }\right] ^{2}\,,
\end{equation*}
\begin{equation}
\eta \equiv 2\,{M_{\rm Pl}^2}\left[ \frac{H^{\,\prime \prime }\left( \phi \right)
}{H\left( \phi \right) }\right]\,. \label{sroll}
\end{equation}
To first order, the slow roll parameters are related to the ratio $r$ of the tensor
to scalar fluctuations, by the relation
\begin{equation} r\sim 16\,\epsilon \end{equation}
and to the spectral index of the scalar $\delta\rho/\rho$ by
\begin{equation}
n_S-1\approx -\frac{3}{8}\,r + 2\eta\,
\label{spind}
\end{equation}
\cite{kolbmay}. The value for $\mu$ that characterizes the end of inflation, is
constrained by the condition that $\epsilon= 1$. From this condition and
Eq.(\ref{Hphi}), we obtain
\begin{equation}
\mu^2= \beta\,. \label{massend}
\end{equation}

Substituting the time at $60$ $e$-folds before the end of inflation from
Eq.(\ref{t60}) and $\beta$ from Eq.(\ref{massend}) into Eq.(\ref{dflut}), we obtain
\begin{equation*}
\frac{\delta \rho }{\rho }=\frac{1}{\sqrt{75}\,\pi M_{\rm Pl}^3}
\,\frac{V^{{3}/{2}}(t)}{V^{\prime}(t){dt}/{d\phi}} \vert_{t=t_{60}}
\end{equation*}
\begin{equation}
\approx 5.42\,\mu  \,.
\label{dflut2}
\end{equation}
Using the above result, together with Eq.(\ref{mast}) and the observational evidence
that the ${\delta \rho }/{\rho }$ produced at $\sim 60$ $e$-folds before the end of
inflation is $ \sim 10^{-5}$, we obtain the predicted value of $M_{\rm end}$, the
mass (energy) scale at the end of inflation,
\begin{equation} M_{\rm end}\approx
10^{13}{\rm GeV}\,.\label{SUSY}\end{equation}

This value is less than the GUT scale ($\sim 10^{14}\,-\,10^{16}{\rm GeV}$), but is
consistent with the upper limits for the mass (energy) scale at the end of inflation
given by Vilenkin \cite{vile2} and Starobinsky \cite{star2}.

Evaluating the spectral index of the scalar $\delta\rho/\rho$ from
Eqs.(\ref{sroll}) and (\ref{Hphi}), we observe that the parameter
$\eta$ is zero and that $\epsilon$ is very small, $\epsilon\simeq
8.3\times 10^{-3}$. From Eq.(\ref{spind}), we have $n_S \simeq
0.95$, an approximately Harrison-Zeldovich spectrum $n_S =1$, in
agreement with the WMAP data \cite{peiris,3WMAP}. These results do not depend on the
exact value of $\phi_0$.

From $\epsilon$ in Eq.(\ref{sroll}) and Eq.(\ref{Hphi}), we obtain
\begin{equation} r=16\,\epsilon \approx 0.13\,.
\label{ratio}
\end{equation}
This value is similar to those predicted by frequently discussed inflation models
with $r\sim 10\%\,-\,30\%$ (e.g., \cite{kolbmay}).

\section{Conclusions}
$\,\,\,\,\,\,\,$ We investigated a model in which the Hubble
parameter is decreasing slowly in time, as predicted  by the
modified Starobinsky model \cite{AIM},\cite{AIMmod},  $H(t)=M_{\rm
Pl}-\beta\,M_{\rm Pl}^2\,t$,  and constructed an inflaton
potential for $H(t)$. The derived potential, normalized at $\sim
60$ $e$-folds before the end of inflation, creates the observed
level of $\delta\rho/\rho\sim 10^{-5}$ and indicates an energy
(mass) scale, $M_{\rm end}\sim 10^{13}\,{\rm GeV}$, at the end of
inflation.

This energy scale at the end of inflation can be compared with those predicted by
Vilenkin and Starobinsky. Vilenkin gave a limiting value $M_{\rm end}\lesssim
10^{16}{\rm GeV}$ for the scale at the end of inflation \cite{vile2}, while
Starobinsky predicted $M_{\rm end}\lesssim 10^{14}{\rm GeV}$ \cite{star2}.

From the inflaton potential, we calculated the spectral index of
the scalar modes. The result, $n_S\simeq 0.95$, is compatible with
the WMAP data. The potential also predicts a tensor contribution,
$r\sim 0.13$, in accordance with most inflation models, which
predict $r\sim 0.10-0.30$ and is in agreement with existing
observational data.

\subsection{Acknowledgments}
R.O. thanks the Brazilian agencies FAPESP (00/06770-2) and CNPq (300414/82-0) for
partial support. A.P. thanks FAPESP for financial support (03/04516-0 and
00/06770-2).



\end{document}